\documentclass[aps,twocolumn,showpacs,floatfix]{revtex4}
\usepackage{amsmath,amsfonts,amssymb,graphicx,color,times,bbm,psfrag,dsfont}
\usepackage[T1]{fontenc}
\usepackage[latin9]{inputenc}

\newcommand{\ket}[1] {\vert{#1}\rangle}

\begin{document}

\title{Gauge-Away Effect in Cold Gases on Optical Lattices}

\author{O. Boada, A. Celi and J. I. Latorre}

\affiliation{
Dept. d'Estructura i Constituents de la Mat\`eria,
Universitat de Barcelona, 647 Diagonal, 08028 Barcelona, Spain
}
\date{\today}

\begin{abstract}
It is shown that a simple modification of the geometry in which Raman lasers are
applied to a cold gas in an optical lattice results in transforming the emerging
effective electromagnetic field into a pure gauge. This contrived {\sl
gauge-away} effect can be observed experimentally by measuring the Mott-Insulator-to-Superfluid critical point. The underlying mechanism for this phenomenon is the ability to
engineer the transfer of the transverse component of the gauge potential into its longitudinal one.
\end{abstract}
\maketitle


Ultracold gases on optical lattices provide a powerful instrument for creating
quantum devices that can simulate, in a controlled manner, a variety of condensed
matter systems. The emerging possibility of acting on the system with carefully
tuned electromagnetic fields invites the design of a new generation of
experiments \cite{LSADSS07}.  In particular, it is in principle possible to
implement effective Abelian \cite{JZ03,M04,SDL05} and non-Abelian \cite{OBSZL05} gauge fields coupled to the Hubbard
model, which describes the physics of an ultracold gas on an optical lattice. For the off-lattice implementations see \cite{JO04,RJOF05}; for the rotation-based approach see \cite{BDZ08}.
This would provide a remarkable realization of man-made effective gauge symmetry.

To obtain a clear understanding of effective gauge fields on optical lattices, simple experiments with neat signatures must be devised to prove the
purported gauge phenomenon. Note that, in this context, the gauge fields are not
real ones, nor are the particles they are coupled to electrically charged
particles. The system is just behaving according to  a Hamiltonian that
simulates an effective, non-dynamic gauge interaction. Therefore, a precise
characterization of what the equivalent of gauge effects are must be put forward.

A word of caution, however, is in order. In the cold atom system the expectation
value of any hermitian operator, in particular the vector potential, is
accessible experimentally. Conversely, in actual gauge systems, be it a quantum
Hall sample or quantum chromodynamics, observables must not only be hermitian but
they must also be gauge-invariant. Since no real dynamic gauge-symmetry is
present in cold atom simulations, the exciting possibility of performing
experiments revealing information that would not be accessible in the real world presents itself.
For instance, gauge-dependent effective vector fields have been observed in  Ref.
\cite{YCPPPS09}, where a Bose-Einstein condensate subjected to an optically
generated effective magnetic field was studied experimentally. The
authors reported distinguishing between different zero magnetic field
configurations. Also, see the viewpoint \cite{J09}.

We here propose a scheme to experimentally simulate a very specific gauge
property of the interactions that are controlled on an optical lattice. The idea
consists in exploiting a simple variation of the geometry of a known experimental
setup to 'gauge-away' the effective, non-dynamic Abelian gauge field. That is,
while no change is made on the intensity of the lasers acting on the optical
lattice, the Abelian  electromagnetic potential they effectively generate can be
transformed at will from a would-be physical field into a would-be pure gauge, i.e. with zero electric and magnetic field. As
a consequence, the measurement of any gauge-invariant observable must reflect the
disappearance of the effective gauge field. Thus, a gauge-away effect could be
observed by means of a simple angle rotation of external laser fields.

Another way of understanding our proposal is as follows. The effective gauge
fields generated in the optical lattices are non-dynamic. The only way to have
a gauge transformation is to modify the action of the external lasers. In a
way, we can physically perform gauge transformations. We can also modify at
will whatever part of a gauge field is longitudinal or transverse and
check how this affects observables. The gauge-away effect corresponds to
specifying the experimental manipulation that produces a transfer from the
transverse to the longitudinal components of a gauge potential.

From a theoretical point of view, our modification of the setup of cold gases on
optical lattices coupled to external magnetic fields needs a detailed analysis.
We shall first present the basic elements of our proposal so as to describe an
experimental signature that would reveal the gauge-away effect. Then, we shall
analyze some of the subtleties that appear in the theoretical description of the
system due to the change in the configuration of the experiment.

Let us begin our discussion by recalling that cold bosonic atoms propagating on a
sufficiently deep optical lattice are adequately described by the single-band
Bose-Hubbard model \cite{JBCGZ98,SCMI02}. In recent years, several proposals have been
put forth whereby the standard setup for generating an optical lattice in the lab is
modified in such a way that the hopping terms in the Hubbard Hamiltonian acquire
a position-dependent phase. For an infinite two-dimensional lattice the
Hamiltonian reads
\begin{eqnarray}\label{hubbard}
H&=&-J_{x}\hspace{-2mm}\sum_{m,n=-\infty}^{\infty}\hspace{-2mm} e^{i\theta^{x}_{m,n}}a^{\dagger}_{m+1,n} a_{m,n}+h.c.-\nonumber \\
&&-J_{y}\hspace{-2mm}\sum_{m,n=-\infty}^{\infty}\hspace{-2mm} e^{i\theta^{y}_{m,n}} a^{\dagger}_{m,n+1}a_{m,n} + h.c.+\\
&&+\frac{U}{2}\hspace{-2mm}\sum_{m,n=-\infty}^{\infty}\hspace{-2mm} N_{m,n}(N_{m,n}-1) - \hspace{-2mm}\sum_{m,n=-\infty}^{\infty}\hspace{-2mm}\mu_{m,n}N_{m,n}\nonumber \, ,
\end{eqnarray}
where $a^{\dagger}_{m,n}$ and $a_{m,n}$ create and destroy an atom at a lattice
site ($m,n$), respectively, and obey the usual bosonic commutation relations. The
constant $J_x$ ($J_y)$ is the site-to-site tunneling energy in the $x$ ($y$)
direction. The parameter $U$ is the pair interaction energy at each site,
$\mu_{m,n}$ is the local chemical potential, and $N_{m,n}$ is the local
occupation. Most remarkably, bosons hopping on the lattice acquire position-dependent
phases $\theta^x(m,n)$ and $\theta^y(m,n)$.

The previous Hamiltonian closely resembles that of charged particles on a lattice
interacting with a classical magnetic field perpendicular to it. As in the case
of ordinary charged particles, from Eq. \ref{hubbard} it is clear that bosons
traveling around a closed loop pick up a phase shift proportional to the magnetic
flux going through the area bounded by the loop. Henceforth, we will identify the
phases $\theta^{i}_{m,n}$, with the components of an effective vector
potential $A^{ef}_i (m,n)$, $i=x,y$. Note that $A^{ef}_i(m,n)$ is only an
effective non-dynamic description of the system. It is not a fundamental field, it does
not propagate, it is not subject to gauge symmetry.

Nevertheless, we can make physical modifications of external lasers that mimic
gauge transformations. This permits the following symmetry. The modified Hubbard
Hamiltonian is invariant under a local transformation of $A^{ef}_{i}(m,n)$ and of the
creation and destruction operators given by
\begin{equation}
\begin{split}
A^{ef}_x(m,n) &\rightarrow A^{ef}_x(m,n) - (\Lambda(m+1,n) - \Lambda(m,n)),\\
A^{ef}_y(m,n) &\rightarrow A^{ef}_y(m,n) - (\Lambda(m,n+1) - \Lambda(m,n)),\\
a_{m,n} &\rightarrow e^{i\Lambda(m,n)}a_{m,n} \, ,\\
a^{\dagger}_{m,n} &\rightarrow e^{-i\Lambda(m,n)}a^{\dagger}_{m,n} \,,
\end{split}
\label{gaugetransformation}
\end{equation}
for any function $\Lambda(m,n)$, which is vividly reminiscent of a gauge
transformation in an ordinary electromagnetic system. Let us stress here that,
despite the invariance of the Hamiltonian under Eq. \ref{gaugetransformation}, in
the cold atom setup the phases $\theta^{i}_{m,n}=A^{ef}_i(m,n)$, and consequently
the effective gauge field, are physical quantities. In order to perform a gauge
transformation on the effective vector potential one must alter the configuration
of the experimental setup, as the local phases $\theta^{i}_{m,n}=A^{ef}_i(m,n)$
must be changed. Note that the invariance of the Hamiltonian under Eq.
\ref{gaugetransformation} maps the solutions of gauge-equivalent configurations
to each other if only gauge-invariant observables are considered. This fact is
far from obvious from the point of view of the standard theory of degenerate
gases.

Let us now present a way to realize the above modified Hubbard model that allows
for an experimental verification of a gauge-away effect. The basic idea remains
to change appropriately the action of external laser fields so as to transfer the
transverse part of the generated effective gauge field into a longitudinal part.
Then, gauge-invariant observables will only depend on the former one.

Of the different ways to create an artificial magnetic field in a two-dimensional
non-rotating optical lattice \cite{JZ03,M04,SDL05}, let us focus our attention on the
experimental scheme proposed in Ref. \cite{JZ03}. In this setup, atoms in
different hyperfine states are trapped in different columns of the lattice by
adequately polarizing the standing wave of light that forms it. The hopping in
the direction of these columns, $x$, is unaltered and as usual it is controlled
only by the lattice depth in that direction, $V_x$. The hopping in the other
direction, $y$, is  controlled by extra lasers which induce Raman transitions
between the two hyperfine states with a complex Rabi frequency. When an atom in a
given hyperfine state changes to the other state, it is no longer at a minimum of
the lattice potential and is compelled to move to one of the neighboring columns.
In order to ensure that the tunneling along $y$ is solely controlled by these
additional lasers, the relation $V_x<<V_y$ must hold. This allows us to set the
non-Raman induced hopping in the $y$ direction, $J_y$, safely to zero in the same
way $J_z$ must be negligible for a two-dimensional optical lattice.

Let $\ket{g}$ and $\ket{e}$ be the two hyperfine levels. The Rabi frequency at
the  lattice position  ${\bf{x}}=(m,n/2)a$  is $\Omega(\textbf{x})=\Omega_0 e^{\pm
i({\bf{k}}_e - {\bf{k}}_g)\bf{x}}$, where ${\bf{k}}_g$ and ${\bf{k}}_e$ are the
wave-vectors of the Raman lasers, $\Omega_0$ is a constant and $a$ is the lattice
spacing. As pointed out in the original work, the tunneling amplitude in the $y$
direction is a complex function and is related to the average  of the Rabi
frequency $\Omega(\textbf{x})$ with Wannier functions, localized and orthogonal superpositions of
Bloch waves, at consecutive sites.
Contrary to the original proposal we will not restrict the vector
$\textbf{q}={\bf{k}}_e - {\bf{k}}_g$ to lie along the $x$ direction. This
modification is what allows the hopping phase in the $y$ direction to depend on
$m$ as well as on $n$,
\begin{equation}
\begin{split}
A^{ef}_y(m,n)\propto (\textbf{k}_e - \textbf{k}_g)\cdot\textbf{x} =m q a\cos\theta + &\underbrace{\frac{n}{2}qa\sin\theta }\,,\\
&\textbf{pure}\, \textbf{gauge}									
\end{split}
\end{equation}
where $q$ is the modulus of $\textbf{q}$ and $\theta$ is the angle between
$\textbf{q}$ and the $x$-axis of the lattice. The proportionality
constant between $A^{ef}_y(m,n)$ and $(\textbf{k}_e -
\textbf{k}_g)\cdot\textbf{x}$ will be computed later on. In this setup, the Raman
transitions induce no translations in the $x$-axis direction, hence  $A^{ef}_x$
is identically zero for all values of $\theta$ and $q$. Moreover the $n$-dependence
of $A^{ef}_y$ is a pure gauge, i.e. it has no effect on the simulated magnetic field
or on any other gauge-invariant observable. That is to say, for $\theta=\pi/2$,
we have a non-trivial, position-dependent hopping phase $A^{ef}_y(n)=\frac{n}{2} q a $,
while the magnetic field is null,
\begin{equation}
B_z=A^{ef}_y(m+1,n) - A^{ef}_y(m,n) \underset{\theta\rightarrow \pi/2}\longrightarrow 0\,.
\end{equation}
Consequently, by rotating the Raman lasers to the special angle $\theta=\pi/2$ the
simulated magnetic field $B_z$ is gauged-away and the system is in effect
simulating a trivial field in a highly non-trivial way.

\begin{figure}
\scalebox{0.8}{\includegraphics{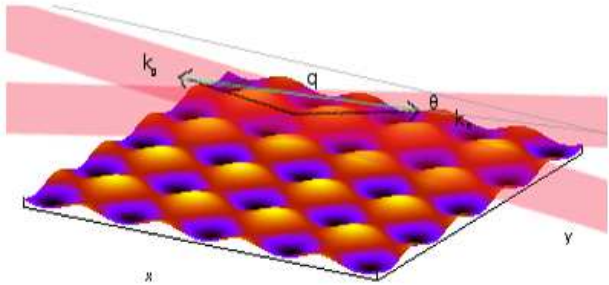}}
\caption{Diagram depicting the geometry of the configuration. The Raman lasers,
with wave vectors ${\bf{k}}_g$ and ${\bf{k}}_e$, are rotated simultaneously with
respect to the lattice in such a way that the angle $\theta$ between the
difference $\bf{q}={\bf{k}}_e -{\bf{k}}_g$ and the $x$-axis, and thus the effective
magnetic flux per plaquette, changes. The pure gauge configuration corresponds to
$\theta=\pi/2$, that is, for $\textbf{q}$ parallel to the $y$-axis.}
\label{diagram}
\end{figure}

From a practical point of view, gauge-invariant observables can be computed in
any gauge. Therefore, we are free to shift $A^{ef}_y \to \bar{A}^{ef}_y$ to a
more convenient gauge. To be precise, we can work out gauge-invariant
observables using the gauge transformation  $\Lambda(n)=qa\sin\theta/4 n(n-1)$
that removes the $n$-dependence of $A^{ef}_y$. Then, for all gauge-invariant
computational purposes, the effective Hamiltonian that governs the bosons on the
optical lattice is
\begin{eqnarray}\label{hubbJZ}
H\hspace{-1mm}&\hspace{-1mm}=\hspace{-1mm}&-J\hspace{-3.5mm}\sum_{m,n=-\infty}^{\infty}
\hspace{-2.5mm} ( a^{\dagger}_{m+1,n} a_{m,n}+e^{i\bar{A}^{ef}_y(m)} a^{\dagger}_{m,n+1}a_{m,n} + h.c.) +\nonumber\\
&&+\frac{U}{2}\hspace{-2mm}\sum_{m,n=-\infty}^{\infty}\hspace{-2mm} N_{m,n}(N_{m,n}-1) -
\hspace{-2mm}\sum_{m,n=-\infty}^{\infty}\hspace{-2mm}\mu_{m,n}N_{m,n} \,,
\end{eqnarray}
where $J$ is calculated by taking the modulus of the average of
$\Omega(\textbf{x})$ with Wannier functions centered at consecutive sites and now
$\bar{A}^{ef}_y(m)=m\alpha\cos\theta$, where $\alpha=\frac{qa}{2\pi}$.

Working with the transformed Hamiltonian, the gauge-away effect can be rephrased.
In the configuration $\theta=\pi/2$, i.e. $\bar{A}^{ef}_y=0$, the system must be
oblivious to the fact that the hopping in the $y$ direction carries the original
associated phase $\theta^y_{m,n}$. In other words, due to the fact that for this
special angle the system is gauge-equivalent to the Bose-Hubbard Hamiltonian with
real, constant hopping amplitudes it follows that the actual system in the lab
must behave accordingly.

To realize the gauge-away effect experimentally it is first necessary to identify
an observable. A natural candidate arises in the context of the Mott insulator
(MI) to superfluid (SF) transition, which has been studied extensively both
theoretically and experimentally in cold atoms on optical lattices (see Ref.
\cite{BDZ08} and references therein). The experimental signature of the MI phase
is the absence of structure (peaks) in the momentum distribution of the gas,
$G(\textbf{k}) \sim \vert w(\textbf{k}) \vert ^2 \sum_{\textbf{R},\textbf{R'}}
e^{ik\cdot(\textbf{R}-\textbf{R'})} \langle
a^{\dagger}_{\textbf{R}}a^{\phantom\dagger}_{\textbf{R'}}\rangle $, where $w(\textbf{k})$ stands
for the Wannier function in momentum space. Conversely, in the SF phase the momentum distribution shows
pronounced peaks at the momentum values of the reciprocal lattice. This is due to
the fact that for the MI phase, $\vert\langle
a^{\dagger}_{\textbf{R}}a^{\phantom\dagger}_{\textbf{R'}}\rangle\vert \rightarrow 0$, while in the
SF phase $\vert\langle a^{\dagger}_{\textbf{R}}a^{\phantom\dagger}_{\textbf{R'}}\rangle\vert
\rightarrow constant$, for $\vert \textbf{R}-\textbf{R'} \vert \rightarrow
\infty$.

A crucial observation is that $\vert\langle
a^{\dagger}_{\textbf{R}}a^{\phantom\dagger}_{\textbf{R'}}\rangle\vert$ at large separation acts as
a gauge-invariant order parameter for the MI-SF transition. Indeed, $\langle
a^{\dagger}_{\textbf{R}}a^{\phantom\dagger}_{\textbf{R'}}\rangle$ transforms multiplicatively with
the phase $e^{i(\Lambda(\textbf{R})-\Lambda(\textbf{R'}))}$ and it only vanishes
in the SF phase. It follows that the MI-SF phase diagram in the $J/U-\alpha$
plain can serve as a witness to the gauge-away effect.

The exact computation of the order parameter for the MI-SF transition is not
possible due to the difficulty of solving the modified Hubbard model exactly for
a large lattice. This observable can be approximated by a mean field computation,
which is known to reflect the MI-SF phase transition. To compute an approximate
ground state of the Hamiltonian in Eq. \ref{hubbJZ} and determine the critical point at
different values of the magnetic flux, we proceed by decoupling the Hilbert
spaces of adjacent lattice sites in occupation space by the replacement
\cite{SKR93} $a^{\dagger}_{m+1,n} a_{m,n}
\rightarrow\psi^{*}_{m,n}a^{\dagger}_{m+1,n}+\psi_{m+1,n}a_{m,n}
-\psi^{*}_{m,n}\psi_{m+1,n}$. The quantities $\psi_{m,n}$, computed according to
the relation $\psi_{m,n}=\langle a^{\dagger}_{m+1,n} \rangle$, are variational
parameters over which the energy is to be minimized. A random initial set of
$\psi_{m,n}$ allows for a diagonalization of the Hamiltonian in the truncated
occupation basis. The eigenvector with lowest lying eigenvalue is then used to
compute a new set of $\psi_{m,n}$. The process is then repeated recursively and
stopped when $\psi_{m,n}$ converges to a fixed point.

\begin{figure}
\scalebox{0.6}{\includegraphics{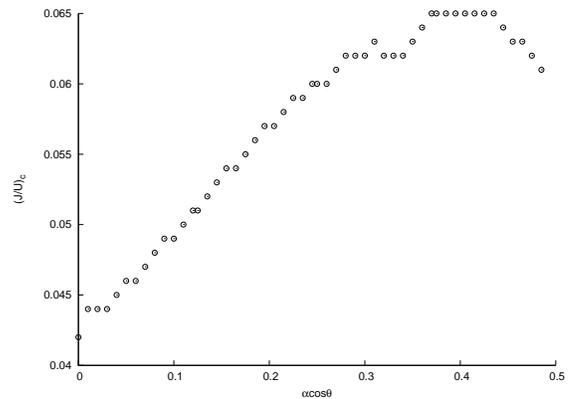}}
\caption{Critical parameter $(J/U)_c$ as a function of the magnetic flux
$\Phi=\alpha \cos\theta$. The phase diagram was determined using the
mean field approximation for a $40\times 40$ square lattice with open boundary
conditions for $\mu_{m,n}=0.5$ and with no trap. As the angle of external lasers is
changed, the transfer from transverse to longitudinal effective gauge field makes
the MI-SF transition to change the location of its critical point. This
corresponds to a gauge-away effect of the effective gauge field.  For $\alpha \cos \theta
\rightarrow 0$, the hopping phase is $\theta^y(m,n)=A^{ef}_y(m,n)=\alpha n/2$,
and the MI-SF transition point appears at the known value $(J/U)_c=.043$.
Although site-dependent phases are present, this transition point remains
unchanged  because $A^{ef}_y(m,n)$  is completely gauged-away, as it only depends
on $n$. Only the interval $\Phi\in\left[0,0.5\right]$ is considered because the Hamiltonian in Eq. \ref{hubbJZ} is invariant
under the replacement $\Phi\rightarrow 1-\Phi$. }
\label{phasediagram}
\end{figure}

The numerical convergence of the iterative mean field method is severely affected
by the addition of the effective gauge field. We find many different minimum
energy solutions of $\psi_{m,n}$. This is consistent with the fact that the
Hamiltonian commutes with the momentum in the $y$ direction and the ground state
is degenerate. In addition, for certain values of $\alpha\cos\theta$ and for certain initial
conditions the process does not converge, it oscillates between two different
non-vanishing values of of $\psi_{m,n}$. Nevertheless, this does not occur in
the MI phase, where $\psi_{m,n}=0$ is the only attractor, so it should not affect the phase
diagram appreciably. The results are consistent with those obtained in a perturbation
expansion in $J$ \cite{NFM99}, and with those obtained by a modified mean field
approach in Ref. \cite{OT07}.

The phase diagram we obtain numerically, see Fig. \ref{phasediagram}, shows an
approximately linear dependence of $(J/U)_c$ on $\alpha\cos\theta$ for a range of
values of $\cos\theta$. As the total effective magnetic flux per unit plaquette,
$\Phi\equiv\alpha\cos\theta$, decreases, the transverse component of the
effective vector potential is transferred to its longitudinal component,
until it is purely longitudinal for $\cos\theta=0$. When this particular
configuration is reached the MI-SF transition takes place at the same value of
$(J/U)_c$ as it does for the un-modified Hubbard model because it is a
gauge-invariant observable that does not depend on the longitudinal part of the
vector potential. That is, the position-dependent hopping phase introduced by the
Raman lasers $\theta^y(m,n)=\alpha n/2$ has no effect (it is gauged away) because
it is identified with the longitudinal component of the the gauge field,
$A^{ef}_y(m,n)=\alpha n/2$. In summary, the gauge away effect we present can then
be experimentally observed as an increase of the point at which the MI-SF phase
transition takes places when the angle of Raman lasers is varied.

Let us now discuss two technical details that have been left apart. First, we
need to check that the modified Hubbard model is a good representation of the
system and, second, the parameters of the model can be fixed at our convenience.
We start by recalling that
the matrix element corresponding to the Rabi
oscillation between the  $\ket{e}$ and $\ket{g}$ states takes the form of an
average over Wannier functions \cite{JZ03}, $J^{ram}_{y}(m,n)=\frac {\hbar}2
\int  \rm d^2{\bf x} \, {\bf w}^*({\bf x}-{\bf x}_{m,n}) \,\Omega\, {\bf w}({\bf
x}-{\bf x}_{m,n+1})$. Using the fact that the Wannier functions factorize, ${\bf
w}({\bf x}) =w_x(x)w_y(y)$, it is
\begin{eqnarray}
&& J^{Ram}_{y}=\frac {\hbar}2 \Omega_0 e^{i\tilde A^{ef}_y(m,n)}\int \rm d \text{$x$}'
\left| \text{$w$}_{\text{$x$}'}(\text{$x$}') \right|^2 \cos( 2 \alpha \cos \theta \,\text{$x$}') \cr
&&\times  \int \rm d \text{$y$}' \text{$w$}^*_{\text{$y$}'}(\text{$y$}') e^{i(2 \alpha \sin \theta \,\text{$y$}')}
\text{$w$}_{y'}(\text{$y$}'- \frac {\pi}2)\, , \label{Jy1}
\end{eqnarray}
where $\tilde A^{ef}_y=2\pi
\alpha(\cos\theta \, m + \sin\theta\, n/2)$. The difference between $\tilde A^{ef}_y$
and the actual gauge field $A^{ef}_y$ is given by a spatially constant (but $\theta$
dependent) phase contribution coming from the integral in the second line and can be omitted.

We are ready to address the first technicality. For Eq. \ref{Jy1} to be a
reliable calculation of the Raman-assisted tunnelling it is necessary that
$\Omega_0$ be small with respect to the optical potential that forms the lattice.
Furthermore, the tilting of the lattice, a crucial ingredient in this
construction, must in turn be small compared to the energy gap to the second band
and be large compared to $\Omega_0$. A numerical evaluation of $J\equiv \vert
J^{Ram}_y \vert$ shows all the above relations can be satisfied simultaneously
for all $\theta$ while meeting the requirement $J=J_x$.

The second technicality is perhaps a more relevant issue. The experimental setup
must remain well calibrated for any value of the magnetic flux $0\le\Phi\le1$.
From Eq. \ref{Jy1} $J=J(\alpha,\theta)$, so the ratio $J/J_x$ could significantly
vary in $\Phi$. If this were the case the degree of anisotropy in the model would
be $\theta$-dependent and could obscure the effects of the effective magnetic
field. For different choices of the lattice potentials $V_{i}$, $i=x,y,z$
compatible with tight-binding approximation and with the requirement that the
free hopping $J_y$ and $J_z$ be suppressed shows that this problem can be avoided
by working at fixed $\alpha$.
For instance, in the case
$V_{0x}=16E_R,V_{0y}=25E_R$, where $E_R$ is the lattice's recoil energy, the
maximum variation $J$ in $\alpha$ is more than $25\%$  while the maximum variation
$J$ in $\theta$ is $4\%$. Note that the two variations go in opposite
directions. It follows that it is possible, at least in principle, to
define the curves $\tilde \alpha(\Phi)$ and $\tilde \theta(\Phi)$ such that $J$
remain precisely constant, $J(\tilde\alpha(\Phi),\tilde\theta(\Phi))=J_x$, $0\le\Phi\le1$.

The following experimental procedure could be used in order to minimize the
deviation of $J/J_x$ from unity. Once suitable values of the $V_{i}$ have been
chosen, we start with a pure gauge configuration (i.e. $\theta=\pi/2$). To
maximize $\alpha$, we fix the relative angle of the  lasers to $\pi$.
At this point, we can perform diffusion experiments as in \cite{SCMI02} for
different Rabi frequencies $\Omega_0$ (whose value is controlled by the
frequencies of the Raman lasers). By determining the value of $\Omega_0$ for
which diffusion rates in the $x$ and $y$ directions become equal, we can, at the
same time, check that the pure gauge configuration corresponds to the free
hopping model with $J(\theta=\pi/2)=J^x$ and prepare the system for the addition
of the effective external magnetic field. Indeed, to turn on $B_z$ we only need
to change the angle $\theta$ from $\pi/2$ without modifying any of the other
parameters.

In conclusion, we have shown that the Hubbard Hamiltonian with position-dependent
complex hopping phases serves as a laboratory to create man-made effective gauge
transformations. It is then possible to devise a gauge-away experiment, based on
the idea of transferring all the transverse part of the effective gauge potential
to  its longitudinal one, only by modifying the angle between the Raman laser
fields and the optical lattice. Thus, a gauge-invariant observable
such as the critical value $(J/U)_c$ for the MI-SF phase transition point changes
in a predictable manner as the effective vector potential is gauged away.

{\bf Acknowledgments}. The authors thank J.I. Cirac and M. Lewenstein for useful discussions and comments. Financial support from QAP (EU), MICINN (Spain),
Grup consolidat (Generalitat de Catalunya),
and QOIT Consolider-Ingenio 2010 is acknowledged.


\begin{thebibliography}{10}

\bibitem{LSADSS07}M. Lewenstein {\it et al.}, Adv. Phys.  {\bf 56}, 243 (2007).

\bibitem{JZ03} D. Jaksch and P. Zoller, New J. Phys. {\bf 5}, 56 (2003).

\bibitem{M04}E.~J. Mueller, Phys. Rev. A {\bf 70}, 041603 (2004).

\bibitem{SDL05}A.~S. S\o{}rensen, E.~Demler  and M.~D. Lukin, Phys. Rev. Lett. {\bf 94}, 086803 (2005)

\bibitem{OBSZL05}K.~Osterloh {\it et al}, Phys. Rev. Lett. {\bf 95}, 010403 (2005).

\bibitem{JO04}G.~Juzeli\=unas and P.~\"Ohberg, Phys. Rev. Lett. {\bf 93}, 033602 (2004).

\bibitem{RJOF05}J.~Ruseckas {\it et al}, Phys. Rev. Lett. {\bf 95}, 010404 (2005).

\bibitem{BDZ08} I. Bloch, J. Dalibard and W. Zwerger, Rev. Mod. Phys. {\bf 80}, 885 (2008).



\bibitem{YCPPPS09} Y.-J. Lin {\it et al.}, Phys. Rev. Lett. {\bf 102}, 130401 (2009).

\bibitem{J09}G.~Juzeli\=unas, Physics {\bf 2}, 25 (2009).


\bibitem{JBCGZ98} D. Jaksch {\it et al.}, Phys. Rev. Lett. {\bf 81}, 3108 (1998).

\bibitem{SCMI02} M.~Greiner {\it et al.}, Nature {\bf 415}, 39 (2002).



\bibitem{SKR93} R.~P. K.~Sheshadri, H. R.~Krishnamurthy and T.~V. Ramakrishnan, Europhys. Lett. {\bf 22}, 257 (2007).

\bibitem{NFM99} M.~Niemeyer, J.~K. Freericks  and H.~Monien, Phys.  Rev. B {\bf 60}, 2357 (1999).

\bibitem{OT07}M.~O. Oktel, M.~Ni\c{t}\u{a}  and B.~Tanatar, Phys. Rev. B {\bf 75}, 045133 (2007).

\bibitem{H76} D.~R. Hofstadter, Phys. Rev. B {\bf 14}, 2239 (1976).


\end{thebibliography}
\end{document}